\documentclass[preprint,superscriptaddress,longbibliography]{revtex4-1}
\usepackage{graphicx}
\usepackage{amsmath,amssymb}
\usepackage{mathrsfs}
\usepackage[normalem]{ulem}
\usepackage{color}
\usepackage{soul}
\usepackage{hyperref}
\usepackage[T1]{fontenc}
\usepackage[utf8]{inputenc}
\usepackage{bm}
\usepackage[left]{lineno}

\usepackage{cancel}

\newcommand{\UFMGfis}{Physics Department, Universidade Federal de Minas Gerais, Belo Horizonte, MG 31270-901, Brazil.}

\newcommand{\UFMGCTNano}{Center of Technology in Nanomaterials and Graphene, Universidade Federal de Minas Gerais, Technological Park of Belo Horizonte, Belo Horizonte, MG, 31270-901, Brazil.}
\newcommand{\UFMGee}{Electrical Engineering Graduate Program, Universidade Federal de Minas Gerais, Belo Horizonte, MG 31270-901, Brazil.}

\newcommand{\NIMS}{National Institute for Materials Science (NIMS), 1-2-1 Sengen, Tsukuba-city, Ibaraki 305-0047, Japan.}
\newcommand{\Boulder}{Department of Physics, and JILA, University of Colorado at Boulder, Boulder, Colorado 80309,
USA}
\newcommand{\vbg}{V_{{\rm BG}}}
\newcommand{\ef}{E_{{\rm F}}}
\newcommand{\mo}{\rm MoS_{2}}
\newcommand{\ws}{\rm WS_{2}}
\newcommand{\wt}{\rm WTe_{2}}
\newcommand{\el}{\rm E_{L}}
\newcommand{\ki}{\rm \chi_{hBN}}
\newcommand{\ti}{\rm \Phi_{Au}}

\begin{document}


\title{Observation of well-defined Kohn-anomaly in high-quality graphene devices at room temperature}

\author{Andreij C. Gadelha}
\thanks{These two authors contributed equally}
\affiliation{\UFMGfis}
\affiliation{\Boulder}
\author{Rafael Nadas}
\thanks{These two authors contributed equally}
\affiliation{\UFMGfis}
\author{Tiago C. Barbosa}
\affiliation{\UFMGfis}
\affiliation{\UFMGCTNano}
\author{Kenji Watanabe}
\affiliation{\NIMS}
\author{Takashi Taniguchi}
\affiliation{\NIMS}
\author{Leonardo C. Campos}
\affiliation{\UFMGfis}
\affiliation{\UFMGCTNano}
\author{Markus B. Raschke}
\affiliation{\Boulder}
\author{Ado Jorio}
\affiliation{\UFMGfis}
\affiliation{\UFMGee}
\date{\today}
\maketitle


{\bf Due to its ultra-thin nature, the study of graphene quantum optoelectronics, like gate-dependent graphene Raman properties, is obscured by interactions with substrates and surroundings. For instance, the use of doped silicon with a capping thermal oxide layer limited the observation to low temperatures of a well-defined Kohn-anomaly behavior, related to the breakdown of the adiabatic Born–Oppenheimer approximation. Here, we design an optoelectronic device consisting of single-layer graphene electrically contacted with thin graphite leads, seated on an atomically flat hexagonal boron nitride (hBN) substrate and gated with an ultra-thin gold (Au) layer. We show that this device is optically transparent, has no background optical peaks and photoluminescence from  the device components, and no generation of laser-induced electrostatic doping (photodoping). This allows for room-temperature gate-dependent Raman spectroscopy effects that have only been observed at cryogenic temperatures so far, above all the Kohn-anomaly phonon energy normalization. The new device architecture by decoupling graphene optoelectronic properties from the substrate effects, allows for the observation of quantum phenomena at room temperature.}

Due to its two dimensional nature, and its  electronic and optoelectronic properties, graphene is a promising material to be integrated in hybrid optoelectronic devices \cite{WANG20191}. For instance, Graphene/hBN heterostructure was used to confine light with small losses \cite{doi:10.1126/science.aar8438}. Besides, a 99.6\% of light absorption was  presented by using a device made of a polymethyl methacrylate (PMMA) grid, graphene, silica, and a gold layer \cite{https://doi.org/10.1002/adom.201600481}. A pursue for alternative devices that detect and collect low-energy photons led Massicotte \textit{et al.} to make an heterostructure based on graphene $\ws$ and hBN that converts low-energy photons into electricity through a photothermoelectric effect \cite{Massicotte2016}. Graphene-based heterostructures were also applied in photoelectric modulators \cite{doi:10.1021/acsphotonics.7b01094}, photodetectors \cite{https://doi.org/10.1002/adma.201706561,Xia2009}, and light-emitting devices \cite{Wu2016}.

Two-dimensional field-effect transistors (2D-FETs) use gate voltages to probe electric field, charge carriers and quantum properties of 2D materials \cite{ben,Cao2021,Pierce2021,Pisana2007,Das2008,PhysRevLett.101.136804,PhysRevLett.98.166802,xubaca}. Especially, the combination of the non-destructive Raman technique, that probes the inelastic scattering of light, with graphene FETs unravels fascinating physics, like the Kohn-anomaly related to the breakdown of adiabatic Born–Oppenheimer approximation, electron-phonon coupling tuning and phonon softening \cite{Pisana2007,PhysRevLett.98.166802,PhysRevLett.101.136804}. However, the theoretically expected tendencies for gated-Raman measurements are not clearly observed at room temperature \cite{Pisana2007}. Fact that is attributed to temperature and charge inhomogeneity in graphene devices \cite{Pisana2007,Das2008,PhysRevLett.98.166802,PhysRevLett.101.136804,PhysRevB.94.075104}. Curiously, such problems was not solved after improvements on device quality, mainly  achieved after incorporating hBN crystals on graphene field-effect devices \cite{bn}, and the main reason is that transparent devices commonly show photodoping effects, which is the name attributed to changes on the density of free charge carriers in 2D FETs attained after light interaction \cite{andreij,Gadelha_2020,pdh}. Here, we present a transparent graphene device that exhibits a clear and straightforward Raman dependency with gate-voltage, free of photodoping effects. We observe the well-defined expected behavior due to the Kohn-anomaly, related to the breakdown of the adiabatic Born–Oppenheimer approximation, at room temperature. Such observations were only achieved because the graphene heterostructures are prepared combining suitable transparent materials to avoid photodoping effects and background optical peaks; improvement that can be extrapolated to  other 2D materials. Our work provides new insights into optoelectronic devices and allows the observation of quantum physics at room temperature with the proposed high-quality graphene device.

To implement field-effect graphene transistor for optoelectronics, graphene is deposited on one side of a insulating material, and another conducting material is placed on the other side to work as gate electrode. In an optical perspective several problems can occur: optical activity of any of the materials involved on the heterostructure can generate background or overlapping signals; the gate material is usually opaque, absorbing the optical-active peaks from graphene; interfaces between materials can be photo-active, impacting the field-effect performance, etc. Below, we describe a device that transcends these limitations. 

\begin{figure*}[!hbtp]
	\centering
	\centerline{\includegraphics[width=183mm]{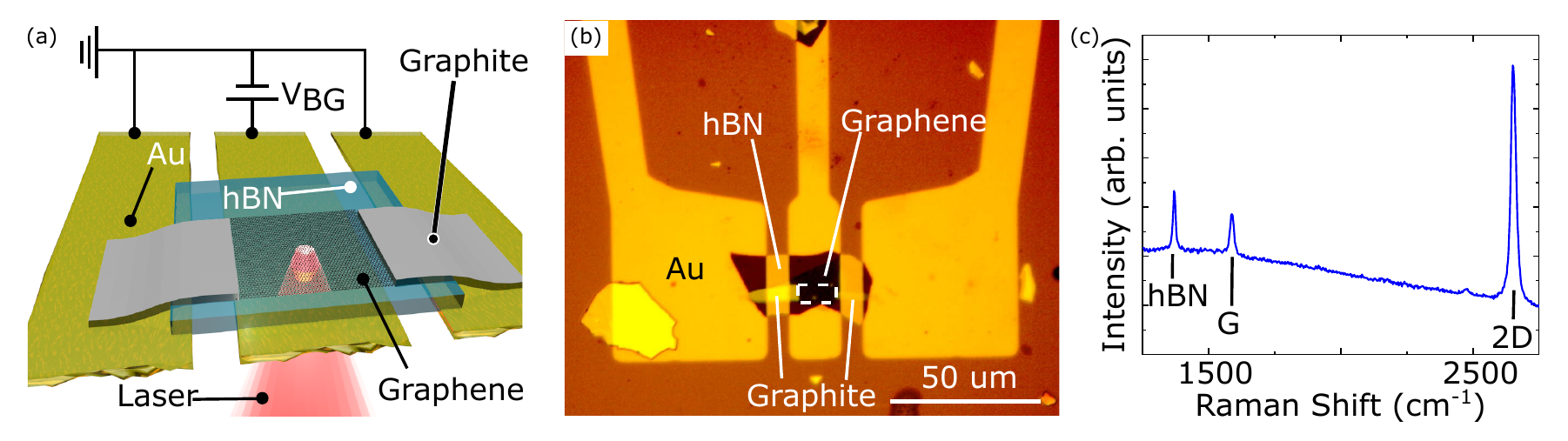}}
	\caption{\label{fig:1}\textbf{Device configuration.} \textbf{(a)} sketch of the device configuration. \textbf{(b)} optical micrograph of the device. \textbf{(c)} Raman spectrum of the Au1 device.}
\end{figure*}

Fig.~\ref{fig:1}(a) shows a sketch of the device used in this work. We first pattern three thin (10~nm) gold (Au) leads by e-beam lithography, then we stack a heterostructure of graphite, graphene, and hBN (in this order) using the pick-up method  and transfer it to the pre-patterned Au leads (see Fig.~\ref{fig:1}(a)). We use the graphite to contact the graphene to the source (S) and drain (D) leads, while we use the gate (G) Au lead to apply a gate voltage ($\vbg$) to graphene. We prepared many devices of this kind, but in this manuscript, we present data of three, which we label Au1, Au2, and Au3. Au1 has a capping hBN layer, and the other devices are not encapsulated. We provide atomic force microscopy (AFM) images of the gold layer that shows a uniform film, see Fig.~S6. Besides, the 10~nm Au layer is thin enough to allow light transmission, being adequate for optical measurements. Fig.~\ref{fig:1}(b) shows an optical image of the device Au1.  Fig.~\ref{fig:1}(c) exhibits a Raman spectrum of this device, where we observe the hBN and the graphene C-C stretching G mode and the second-order breathing 2D mode Raman peaks. Although we measure a small background from the 10 nm think Au layer, the graphene peaks are clearly observed, see Fig.~\ref{fig:1}(c). Hence, the device presented in Fig.~\ref{fig:1} exhibits transparency and the absence of significant background optical peaks and luminescence. For completeness, we present investigations with devices using $\ws$, $\mo$, and $\wt$ gate electrodes which do not show better results than the Au, see Figs.~S1 and S2, .

Photodoping effects are typically observed when transistors made of 2D materials are exposed to light exposure \cite{fepn,pdh,andreij,Gadelha_2020}. Although a photodoping effect permits applications on 2D photomemory devices \cite{andreij}, it is a critical problem for high-performance optoelectronic devices based on 2D-FETs. 
On a 2D-FET, the h-BN crystals work as insulators between the graphene and gate. Because hexagonal boron nitride crystal has an energy gap of $\sim$ 5.9 eV \cite{BHIMANAPATI2016101}, we do no expect any optoelectronic activity with the visible light. However, hBN loses its insulating property under interaction with visible light in some 2D-FETs. The reason is that the visible light photons excite the electronic states at the interface between hBN and the gate, breaking hBN insulating performance. Thus, the field-effect performance is affected, and charges are trapped in the heterostructure giving rise to photodoping \cite{andreij}. 

\begin{figure*}[!hbtp]
	\centering
	\centerline{\includegraphics[width=89mm]{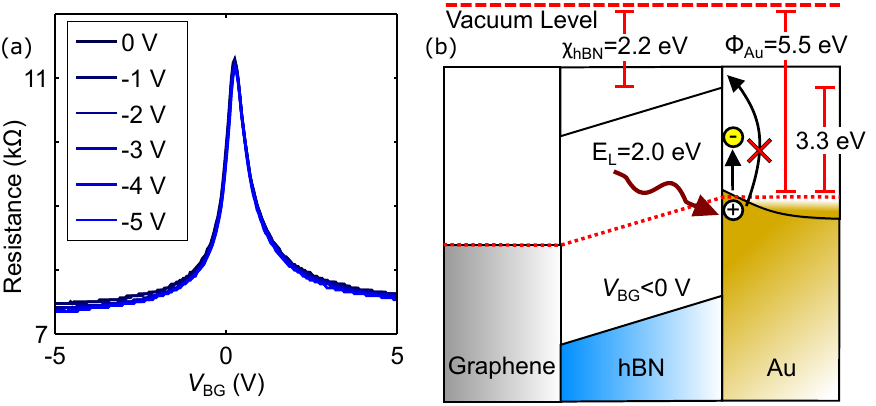}}
	\caption{\label{fig:4}\textbf{photodoping generation in Au devices.} \textbf{(a)} R vs $\vbg$ curves measured after laser exposures with the applied gate shown on legend. Exposure time of two minutes and laser power of 200 $\mu$W. \textbf{(b)} photodoping model for the graphene-hBN-Au junction.}
\end{figure*}

So, the photodoping hinders a proper investigation of the intrinsic properties of graphene, as well as its field-effect performance. For instance, reference \cite{Pisana2007} does not provide a data for G frequency and FWHM as a function of $\vbg$ that reveals clear Kohn-anomalies and electron-phonon coupling results. They use silicon with thermally grown SiO$_{2}$ layer as substrate which is known to present photodoping \cite{fepn,Gadelha_2020}. Fig.~\ref{fig:4}(a) shows Resistance (R) as a function of $\vbg$ curves for the Au1 device to inspect its possible photodoping generation. For each curve, we keep $\vbg$ at each potential and expose the device to the laser for two minutes. Then, we turn the laser off and repeat the R vs $\vbg$ curve. This procedure was used to investigate photodoping, translated as shifts of the charge neutrality point (CNP) \cite{fepn,pdh,andreij,Gadelha_2020}. 

 Because there are no CNP shifts for all exposures in Fig~\ref{fig:4}(a), our device induces no photodoping in graphene. Thus, in such device the gate-voltage application does not compromise the measurement of graphene optoelectronic properties. We present in Fig~\ref{fig:4}(b) the energy diagram of the graphene/hBN/Au junction as proposed in reference \cite{andreij} to understand why our device has no photodoping. For negative gate voltages photoexcited electrons from Au tend to flow towards graphene. However, this process only occurs if the laser energy ($\el$) is larger than the difference between Au work function ($\ti$) and hBN electron affinity ($\ki$). Thus, we expect that metals with higher work function will generate lower photodoping. 
Au work function is $\ti=$5.5~eV \cite{Holzl1979}, which is the largest in this work, making Au a less photodoping gate material. Recall that the calculated work functions for WS$_{2}$, MoS$_{2}$ and WTe$_{2}$ are approximately 4.5~eV, 4.75~eV 4~eV \cite{PhysRevB.103.085404}. In Fig~\ref{fig:4}(b), $\ti-\ki=\mathrm{3.3\,eV}$, while $\el=\mathrm{2.0\,eV}$, which point out for the absence of photodoping generation in the proposed device. The reason to use a capping hBN on Au1 device in Fig~\ref{fig:4}(b) measurement is to avoid laser-induced doping from air-contaminants. We also provide in Fig.~S5 a leakage current test for Au1 device.

\begin{figure*}[!hbtp]
	\centering
	\centerline{\includegraphics[width=183mm]{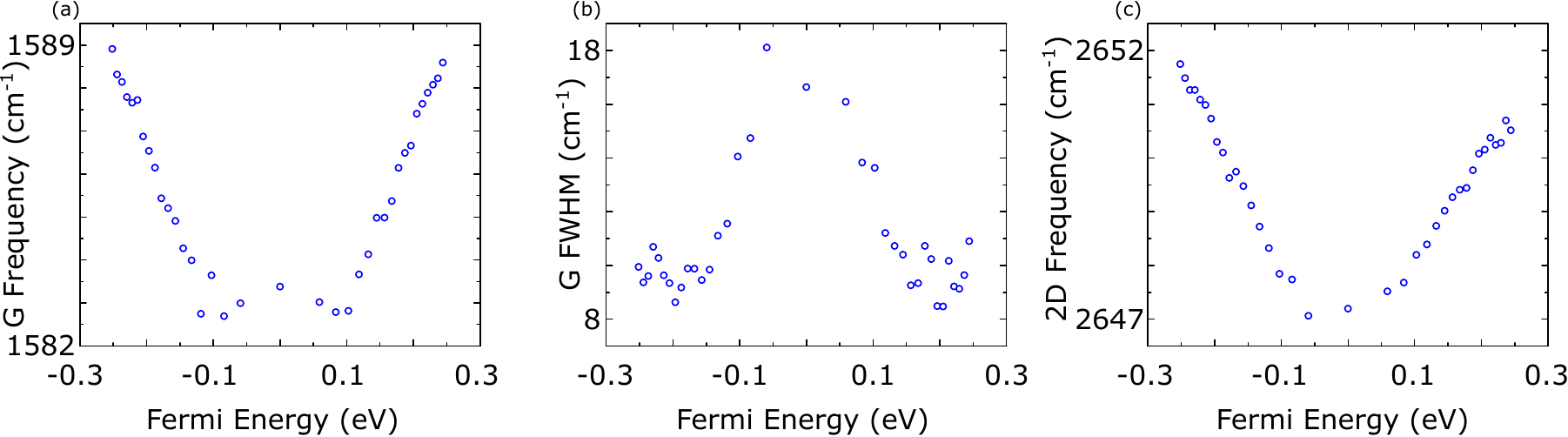}}
	\caption{\label{fig:2}\textbf{Raman feature properties as a function of the graphene Fermi level.} \textbf{(a)} and \textbf{(b)} frequency and FWHM as a function of Fermi energy for the G band, respectively. \textbf{(c)} frequency as a function of Fermi level for the 2D band. This measurements are from the Au2 device.}
\end{figure*}

To test the quality of the Au optoelectronic device, we analyse the frequency and full-width at half maximun (FWHM) of the G mode, see Fig.~\ref{fig:2}(a-b), and the frequency of the 2D mode (see Fig.~\ref{fig:2}(c)) as a function of Fermi-level for the Au2 device. We plot Fermi-level ($E_{\mathrm{F}}$) instead of $\vbg$ to provide a better picture on the graphene's physical properties. Besides, we only plot the G FWHM and G and 2D frequencies on Fig.~\ref{fig:2} because they have well-know physical mechanisms \cite{Pisana2007}. We evaluate the Fermi-level from the $\vbg$ using the formula:

\begin{equation}
\label{eq1}
E_{\mathrm{F}}=\hslash v_{\mathrm{F}}\sqrt{\pi\alpha\left(\vbg-V_{\mathrm{CNP}}\right)},    
\end{equation}

\begin{equation}
\label{eq2}
\alpha=\frac{\epsilon_{0}\epsilon_{\mathrm{hBN}}}{ed},    
\end{equation} where $v_{\mathrm{F}}=10^{6}\,\mathrm{m/s}$ is the graphene's Fermi velocity \cite{Hwang2012}, $V_{\mathrm{CNP}}=0.2\,\mathrm{V}$ is the potential at graphene's charge neutrality point, $\epsilon_{\mathrm{hBN}}=3.5$ is the hBN dielectric constant \cite{hBNd}, and $d=30\,\mathrm{nm}$ is the hBN thickness.

Fig.~\ref{fig:2}(a) exhibits an expected modulation of the G frequency when changing $\ef$. Interestingly, for $\left|\ef\right|\approx 0.1\,\mathrm{eV}$ we observe two dips on the G frequency, which clearly show theoretically predicted details of the Kohn anomaly. The energy difference of these two dips is $\approx 0.2\,\mathrm{eV}$, as defined by the G phonon energy (\cite{Lazzeri2006}). Fig.~\ref{fig:2}(a) shows the raw data, with no treatment and no guide to the eyes nor modeling. Fig.~\ref{fig:2}(b) shows a peak of G FHWM around 0~eV with a width of $\approx 0.2\,\mathrm{eV}$, again clearly delimitating the G phonon energy. For $\left|\ef\right|\geq0.1\,\mathrm{eV}$, G phonons cannot generate electron-hole pairs, increasing its lifetime and decreasing its energy linewidth \cite{PhysRevLett.98.166802} (approximately, since this is only absolute for T = 0\,K). For completeness, Fig.~\ref{fig:2}(c) presents the 2D frequency vs $\ef$. Here the Kohn-anomaly is not so well-defined, as expected, since the 2D peaks is related to phonos out of the high symmetry K-point, where the Kohn-anomaly is well marked. We provide similar data for the Au1 and Au3 devices in Figs~S2 and S3. Therefore, Fig.~\ref{fig:2} provides unprecedented data with features that have only been experimentally observed so far at cryogenic temperature. Such clear data is achieved because the graphene FET is composed suitable materials to be ``gateble'' and photo-gating free.

In summary, we studied a high-quality graphene device, made of hBN and a thin Au gate. We showed that this device allowed deep and clear data from the G Raman peak frequency and FWHM as a function of Fermi level. Accordingly, we observed the well-defined W-shaped Kohn-anomaly effects, related to the breakdown of adiabatic Born–Oppenheimer approximation, at room temperature. We attributed this achievement to our device aspects, which included transparency, absence of optical peaks close to G and 2D peaks, and lack of photodoping. Our work points toward the easier observation of graphene and other 2D materials intrinsic properties by using a high-quality and specific optoelectronic device.
. 


\subsection*{References}
\bibliographystyle{unsrt}

\subsection*{Acknowledgments}  
This work was supported by CNPq (302775/2018-8, 436381/2018-4, 305881/2019-1, and INCT/Nanomaterials de Carbono), Finep (SibratecNano 21040), CAPES (RELAII and 88881.198744/2018-01) and FAPEMIG, Brazil. A.C.G. acknowledges partial support from DoE Award No. DE-SC0008807. 

\subsection*{Authors contributions}  

{\bf Sample preparation:} Andreij C. Gadelha and Tiago C. Barbosa. 
{\bf Raman measurements:} Andreij C. Gadelha and Rafael Nadas.
{\bf Data Analysis:} Andreij C. Gadelha, Rafael Nadas and Ado Jorio.
{\bf Project idealization and guidance:} Ado Jorio, Andreij C. Gadelha, Leonardo C. Campos and Markus B. Raschke. 
{\bf Paper writing:} Ado Jorio, Andreij C. Gadelha and Rafael Nadas.

\subsection*{Author Information} 

The authors declare no competing financial interests. Correspondence and requests for materials should be addressed to Ado Jorio (adojorio@fisica.ufmg.br).

\subsection*{Data Availability Statement} 
The experimental data related to this work can be obtained upon request to the contact authors.

\end{document}